# Molecular Dynamics Study to Predict Thermo-Mechanical Properties of DGEBF/DETDA Epoxy as a Function of Crosslinking Density

Sagar Umesh Patil[1], Sagar P. Shah[2], Michael N. Olaya[2], Prathamesh P. Deshpande[1], Marianna Maiaru[2], Gregory M. Odegard[1*]

[1]Michigan Technological University, Houghton, MI-49931, USA

[2]University of Massachusetts, Lowell, MA-01854, USA

**ABSTRACT**

Epoxy resins are used extensively in composite materials for a wide range of engineering applications, including structural components of aircraft and spacecraft. The processing of fiber-reinforced epoxy composite structures requires carefully-selected heating and cooling cycles to fully cure the resin and form strong crosslinked networks. To fully optimize the processing parameters for effective epoxy monomer crosslinking and final product integrity, the evolution of mechanical properties of epoxies during processing must be comprehensively understood. Because the full experimental characterization of these properties as a function of degree of cure is difficult and time-consuming, efficient computational predictive tools are needed. The objective of this research is to develop an experimentally-validated Molecular Dynamics (MD) modeling method, which incorporates a reactive force field, to accurately predict the thermo-mechanical properties of an epoxy resin as a function of the degree of cure. Experimental rheometric and mechanical testing are used to validate an MD model which is subsequently used to predict mass density, shrinkage, elastic properties, and yield strength as a function of the degree of cure. The results indicate that each of the physical and mechanical properties evolve uniquely during the crosslinking process. These results are important for future processing modeling efforts.

## 1. Introduction

Thermosetting epoxies are an excellent candidate to be utilized for variety of applications in the aerospace industry. They possess excellent mechanical, thermal, and electrical properties[1]. In addition, epoxies are widely-used for benchmarking research as they are inexpensive, and epoxy-based composites are relatively easy to fabricate following simple curing protocols[2]. During the curing of composite panels, the epoxy matrix is subjected to elevated temperatures and the corresponding thermo-mechanical properties evolve as the curing progresses. Additionally, the epoxy matrix experiences shrinkage during the cure process. Because the evolving matrix thermo-mechanical properties and shrinkage drives the development of residual stresses in composite panels during the curing process, a better understanding of the evolution of these properties is important for optimizing the composite processing parameters. The full experimental characterization of composite property/shrinkage evolution is time-consuming and expensive.

Molecular dynamics (MD) is a powerful tool to simulate molecular behavior and provide insight into the effect of temperature, pressure, and other design parameters on the complex networked molecular structure that epoxies possess[3, 4]. As a result, these simulations can also provide atomistically-informed predictions of the evolution of thermo-mechanical properties and shrinkage of epoxy during processing. All MD techniques utilize a force field to describe the interaction of bonded and non-bonded atoms in the molecule. The correct choice of force field based on the design objectives is necessary to accurately simulate molecular behavior and predict accurate properties.

For decades, MD simulations of polymer-based engineering materials have been mostly limited to using fixed-bond force fields with harmonic bonds, that is, force fields that do not simulate the formation or scission of covalent bonds. Specifically, MD modeling studies for neat epoxies have employed various fixed-bond force fields such as OPLS-UA [5, 6], CHARM and cff91 [7], CVFF [8], Dreiding [9-13], and COMPASS [14]. Reactive force fields allow for the direct simulation of the formation and/or scission of covalent bonds during the simulated deformation process, thus allowing for accurate predictions of strength and mechanical properties at large deformations. In particular, the reactive force field ReaxFF has been used to simulate epoxy systems and accurately predict mechanical properties[15-17]. However, ReaxFF is highly complex, thus MD simulation times utilizing this force field are prohibitively long and limited to relatively small simulations.

Recently, Winetrout et al.[18] developed the Reactive Interface Force Field (IFF-R) which combines the efficiency of a fixed-bond forced field with the capability of simulating bond scission. IFF-R replaces the traditional harmonic bond-stretching potential with a morse potential to model covalent bond dissociation in response to large local mechanical deformations. Odegard et al.[19] demonstrated the capability of IFF-R for accurately predicting mechanical properties of fully-cured epoxy systems, however, IFF-R has not yet been used to simulate partially-cured epoxy systems.

All of the MD modeling studies of epoxy cited above provide predicted properties for either a fully cured epoxy or a semi-cured epoxy at a limited number of intermediate crosslinking densities. None of these studies provide predicted properties for a large range of crosslink densities, which is necessary for comprehensive, multiscale process modeling methods for optimizing thermo-mechanical properties of epoxy composites and minimizing process-induced residual stresses[20, 21].

This work utilizes MD simulation to accurately predict the physical and mechanical properties of an epoxy system at varying degrees of cure ranging from fully uncrosslinked to fully crosslinked states at room temperature (300 K, 27 °C). This study implements IFF-R to accurately predict physical and mechanical properties corresponding to large deformations. The modeling approach used herein is validated using experimental characterization of the epoxy system. The results of this study provide the input necessary for comprehensive process modeling of epoxy composites, and key insight into the gelation process of epoxy during curing.

## 2. Molecular Modeling

### 2.1 Material System and Force Field

For this study, an epoxy system comprised of diglycidyl ether bisphenol F (DGEBF) resin and diethyltoluenediamine (DETDA) hardener was modeled using the IFF-R force field. Figure 1 shows the molecular structures of both monomers. The LAMMPS[22] (Large-scale Atomic/Molecular Massively Parallel Simulator) software package was used to perform all the MD simulation for this work. The properties predicted are the bulk mass density ($\rho$), post-gelation volumetric shrinkage ($\Delta V/V$), bulk modulus ($K$), shear modulus ($G$), Young's modulus ($E$), Poisson's ratio ($v$), and yield strength ($\sigma$).

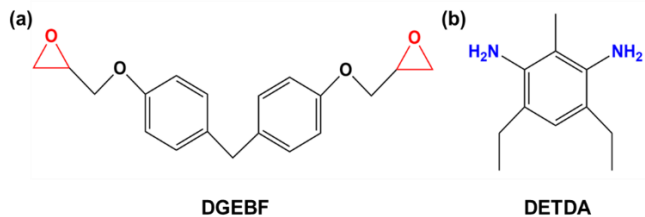

**Figure 1.** Molecular structure of (a) DGEBF (EPON 862) resin and (b) DETDA (Epikure W) hardener. The reactive epoxide groups are shown in red color and diamine groups in blue color.

## *2.2 Model Setup*

For the first step in the MD modeling procedure, a series of equilibrated MD models were constructed. First, both the monomers were assigned IFF-R parameters and combined in a large orthogonal simulation box with a resin-to-hardener ratio of 2:1. The system was replicated to produce 90 monomers of DGEBF and 45 monomers of DETDA to create a low-density MD model containing 5265 atoms. To allow the monomers to mix, a fixed-volume and fixed-temperature (NVT) simulation was performed over 100 picoseconds (ps) with 1 femtosecond (fs) time steps at 600 K.. Following this mixing simulation, the simulation box was slowly compressed (densified) to a target density of 1.17 gm/cc. This simulation occurred at 300 K over 8 nanoseconds (ns) at 1 fs time steps. After densification, an annealing simulation was performed, during which the temperature was ramped from 300 K to 600 K and then slowly cooled to 300K at a 20K/ns cooling rate with the NVT ensemble. A fixed-pressure and fixed-temperature (NPT) simulation was performed at 300 K and 1 atm for 1 ns with 1 fs timesteps to allow the atoms to reconfigure and attain stable energy positions. By repeating this procedure, five different replicates were created to account for statistical deviations in predicted properties. The Nose-Hoover thermostat and barostat were implemented for all the simulations discussed herein[23-25]. Once all the 5 replicates were annealed and relaxed, the next step was crosslinking. Figure 2 shows a representative MD simulation box of a fully densified and equilibrated system.

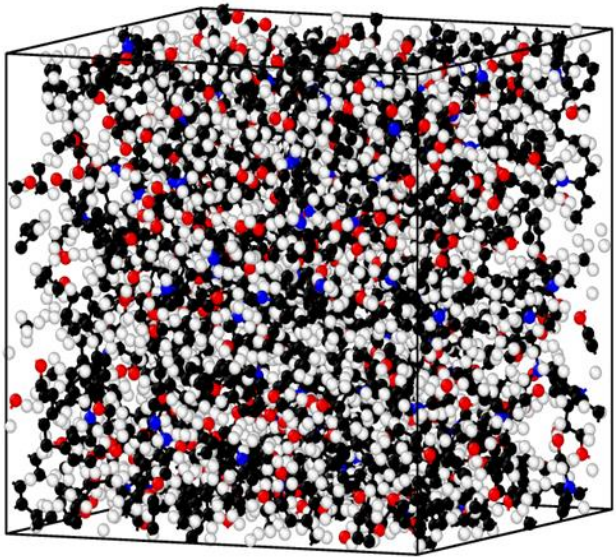

**Figure 2.** MD model consisting of densified and equilibrated epoxy system.

## *2.3 Crosslinking of epoxy monomers*

Figure 3 shows the two-step crosslinking reaction in the DGEBF/DETDA epoxy system. In the first step, the epoxide ring (red color) opens, and the exposed epoxide carbon reacts with a nitrogen from a primary amine (blue color in step 1) of DETDA and forms a hydroxyl group (purple color) and a secondary amine (blue color in step 2). In the second step, the newly formed secondary amine reacts with the epoxide carbon of another DGEBF monomer to form a tertiary amine (blue color) and a second hydroxyl group (purple color). The "fix bond/react" command developed by Gissinger et al.[26] in LAMMPS was implemented to simulate these reactions. This command allows for the simulation of user-defined crosslinks. The pre-reaction, post-reaction, and mapping templates were created accordingly. These simulations were performed at 450 K with 0.1 fs timesteps. Individual models with crosslinking densities varying from 0 to 0.95 were created for all 5 replicates. Here, the crosslinking density ($\phi$) is defined as the ratio of number of crosslinks formed to the maximum possible crosslinks that could be formed in the entire system. During these simulations, the density and volume of the simulation boxes were tracked. Following the crosslinking, an NPT simulation was performed for each crosslinking density at 300 K for 1 ns at 1 fs timesteps to relax the model and to predict the final density and volume. The volumetric shrinkage was calculated as the percent change in the volume of the crosslinked model at specific crosslinking densities, with respect to the uncrosslinked model. For all simulations, the Nose-Hoover barostat was set to maintain a 0.101 MPa (1 atm) pressure on all sides of the simulation box.

Two crosslinking strategies were investigated as shown in Figure 4. For the non-sequential approach, each crosslinked model was generated starting from the same uncrosslinked ($\phi = 0$) model. That is, crosslinking steps were performed directly from 0 to 0.1, 0 to 0.2, and so on, up to the last step, 0 to 0.95. For the sequential approach, each crosslink density was achieved from the increment before it. That is, crosslinking was performed from 0 to 0.1, then 0.1 to 0.2, and so on up until the final increment of 0.9 to 0.95. The progressive building of crosslinks in the sequential strategy mimics the actual curing of polymers and hence was adopted for this study, although a comparison of predicted properties from both approaches is included in the supplementary information.

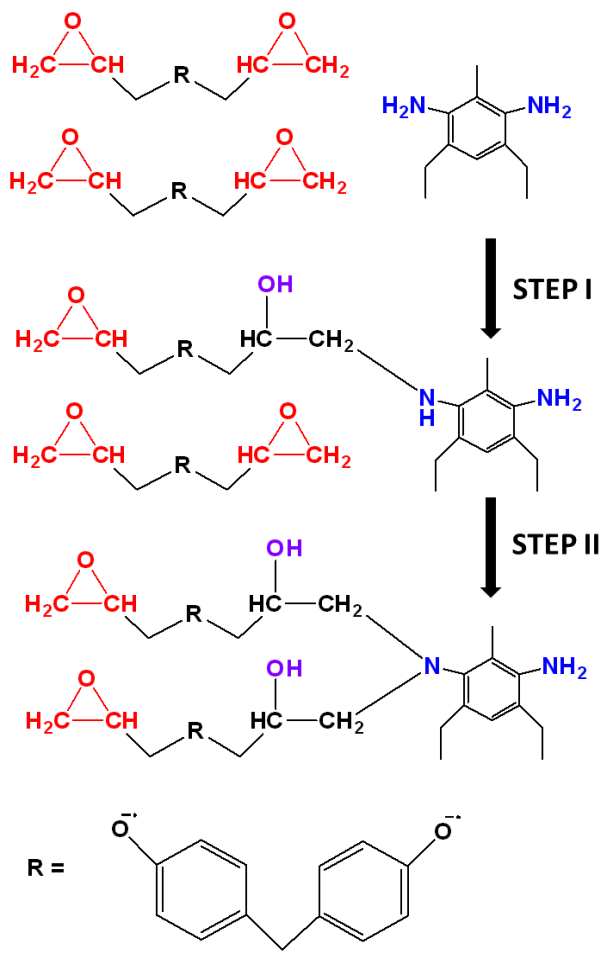

**Figure 3.** Epoxide-amine crosslinking reaction in DGEBF/DETDA epoxy system.

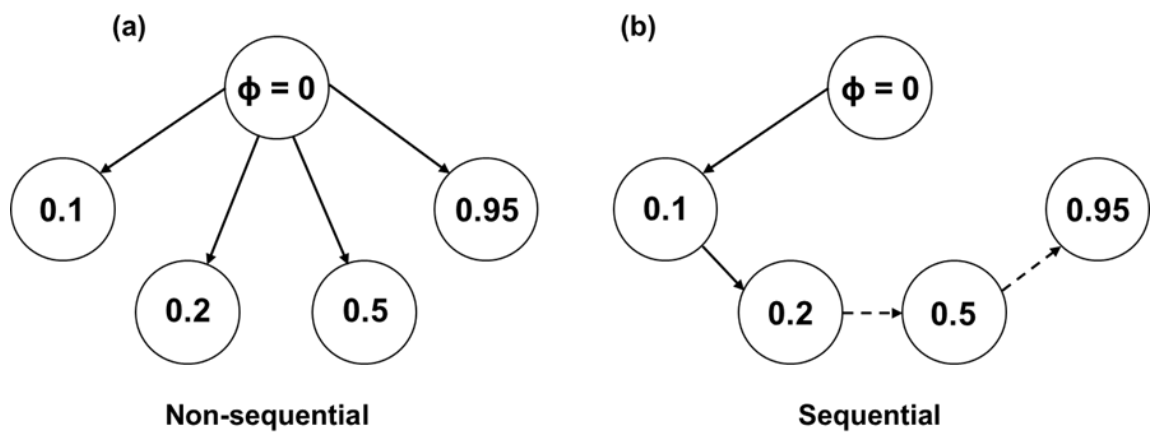

**Figure 4.** Crosslinking strategies (a) non-sequential, (b) sequential

## *2.4 Gel point prediction*

According to Flory[27], an infinite network is formed at the polymer gel point where a single molecule is large enough to span the entire length of the system. Here, the largest molecule represents the largest cluster in the simulation box wherein all the resin and hardener monomers are connected. The gel point can be predicted via MD simulation by tracking the molecular weight of the largest cluster, the molecular weight of the second-largest cluster, or the weight-averaged reduced molecular weight (RMW) of the system [8, 28, 29]. The RMW is the molecular weight of the entire system except the largest cluster. When using the largest cluster metric, the inflection point at which the largest cluster molecular weight drastically increases marks the gel point. When using the other two metrics, the gel point occurs at the point where the corresponding molecular weight reaches a peak value.

In this work, the gel point was predicted using all three methods. Figure 6 shows representative results for all three metrics for one replicate. The average gel points of the five replicates were 0.63, 0.60, and 0.62 for the largest cluster, second-largest cluster, and RMW metrics, respectively. These values agree well with other modeling results from the literature for the same epoxy sytem[8, 28, 29]. Figure 7 shows representative snapshots of the largest cluster at different crosslinking densities, demonstrating the evolution of an infinite network in a periodic simulation box. The "Cluster analysis" modifier in OVITO[30] was implemented to generate these images. When the crosslinking begins, the simulation box contains only monomers. As the crosslinking progresses, covalent bonds form between the monomers and the network starts to grow. In Figure 7, at 0.2 crosslinking density, the network formation initiates. At a crosslink density of 0.4, the network is not yet large enough to span the entire simulation box. Finally, at 0.7 the largest cluster spans the simulation box, which indicates that a continuous load-carrying covalent bond network exists in each direction, and the polymer has reached the gel point. The gel point generally varies between replicates and is highly dependent on the starting molecular configuration. The value of $\phi = 0.6$ was chosen as the gel point of this system for calculation of the post-gelation volumetric shrinkage because it is the simulated value that is closest to the average gel points predicted using all three metrices.

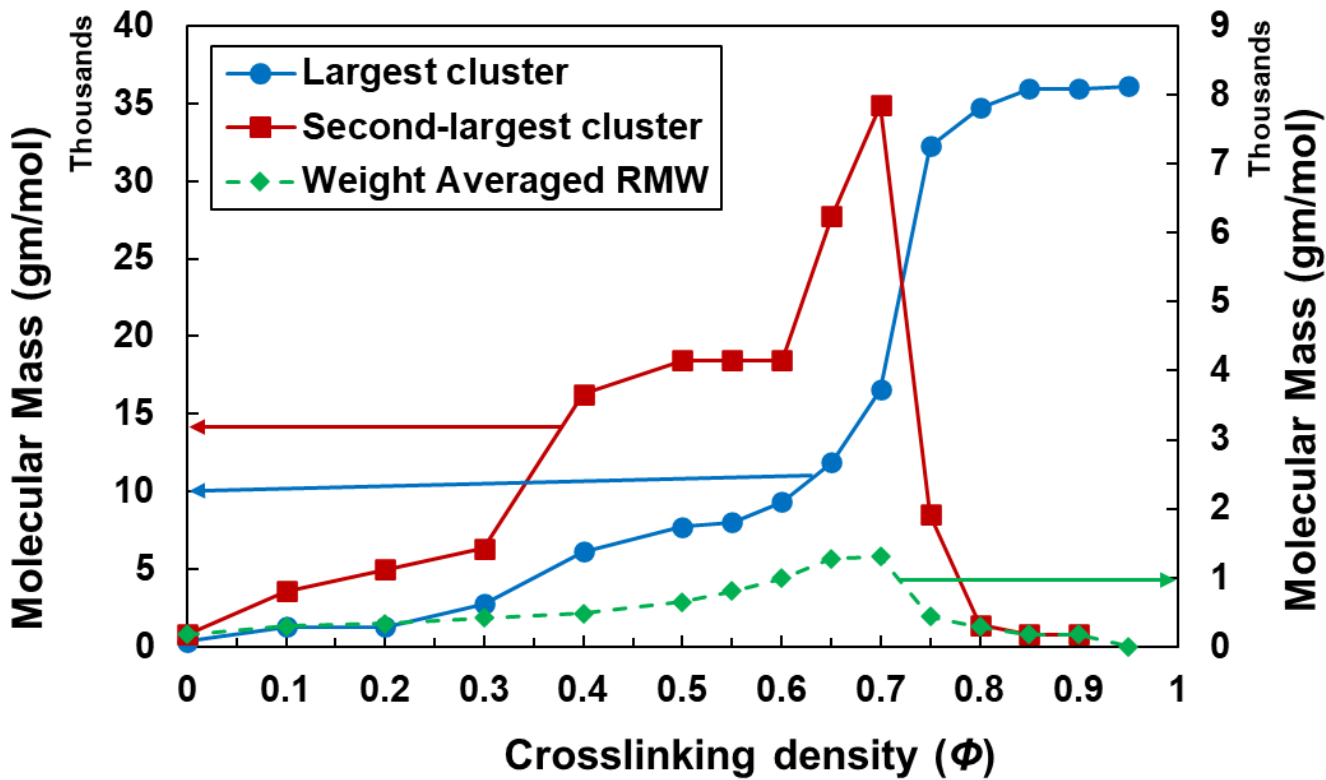

**Figure 6.** Molecular mass versus crosslinking density for a representative system.

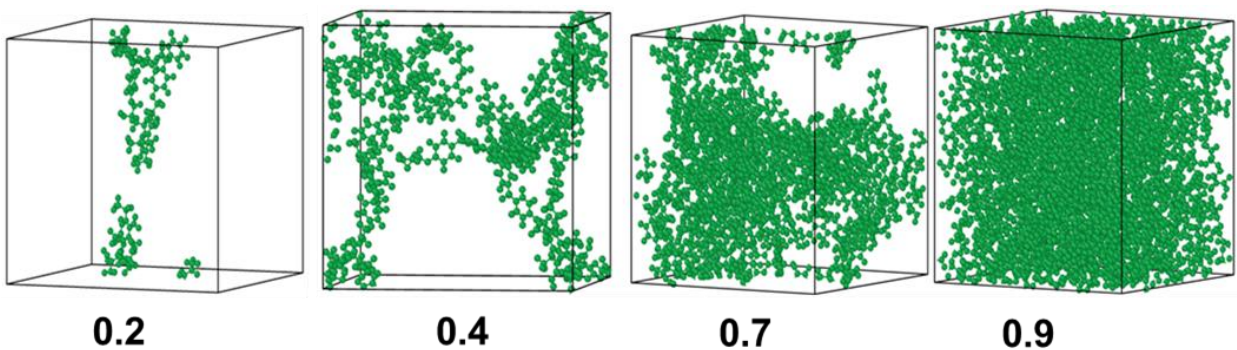

**Figure 7.** Representative snapshots of largest cluster at varying crosslinking densities.

## *2.5 Mechanical deformations and thermal analysis*

The next step in the analysis was to perform simulated mechanical deformations. Here, $K$, $G$, $E$, $v$ and $\sigma$ for each crosslinking density for all replicates were predicted by implementing the similar simulation procedures as outlined by Odegard et al[19]. For each replicate and crosslink density, $K$ was predicted from 1 atm and 5000 atm NPT simulations at 300K, and $G$ was determined from shear deformations in the three principal planes. The ($K$, $G$) pair was then used to predict the ($E$, $v$) pair using the standard relations for linear elastic isotropic materials[31]. The yield strength was predicted using the von Mises stress calculated from the shear deformations [].

## 3. Experimental details

The experimental procedure for the measurement of the mass density is described by Odegard et al[19]. This section describes the measurement of gel point, volumetric shrinkage, and mechanical properties for the DGEBF/DETDA epoxy.

### *3.1 Specimen Preparation*

The DGEBF/DETDA epoxy resin system components were first measured according to a stoichiometric ratio of 100:26.4 (parts by weight) and then mixed thoroughly by hand for two minutes at room temperature. Significant air entrainment as a result of the mixing process was observed as is evident in Figure 8a. To eliminate the trapped air, the uncured resin mixture was first heated in an oven held at 80C for 20 minutes, then degassed at room temperature for an additional 20 minutes. The heating step served as means to reduce the resin viscosity for a more effective degassing procedure. The reduced resin viscosity allowed for a more effective degassing procedure. The degassing process and void-free resin is shown in Figure 8b. For the gel point and volumetric shrinkage characterization described below, the degassed mixture was directly tested in this state. For the mechanical characterization, the degassed mixture was injected into an open-faced mold for full curing of tensile test specimens. The specimens were manufactured according to ASTM D638 Type I specifications. Once fully cured and cooled to room temperature, each specimen was water-polished to eliminate any irregularities found in the cross-sectional area along the gage length of the specimen as a result of curing in open-faced molds. After polishing, a black/white contrasting speckle pattern was applied to the specimens in preparation for 2-D Digital Image Correlation (DIC) strain measurement[32, 33] during tensile testing.

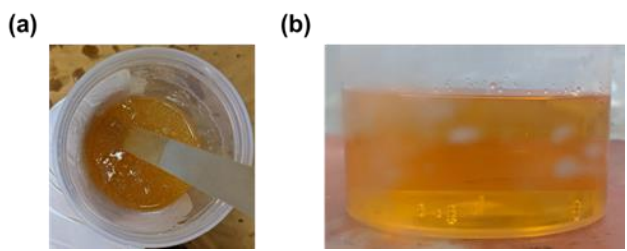

**Figure 8.** (a) Significant air entrainment present after hand mixing DGEBF/DETDA. (b) Degassing resin after pre-heating to remove trapped air.

### *3.2 Gel point and volumetric shrinkage testing*

In order to experimentally evaluate the gel point and the post-gelation chemical shrinkage ($\varepsilon_{sh}$) of the DGEBF/DETDA epoxy, a rotational rheometer (ARES-G2, TA Instruments) with a parallel plate setup was used. The resin mixture, prepared following the aforementioned procedure, was injected between the parallel plates which had an initial gap $h_0 = 1.5 \pm 0.05$ mm. The mixture was then rapidly heated to the desired temperature and allowed to soak under isothermal conditions. For this study, the tests were conducted at two isothermal temperatures of 150 ºC and 170 ºC.

Gel point measurements were carried out in gap-control mode. As the resin mixture cured, it was subjected to an oscillating shear strain which induced a shear stress in the curing resin. The complex shear modulus $G$, decomposed into its elastic storage shear modulus $G'$ and viscous loss shear modulus $G''$, was measured by the rheometer. The point during the cure process at which both $G'$ and $G''$ intersect (have equal values) indicates the transition of the material from a primarily liquid/viscous to solid/elastic phase. This point was used to determine the time to gelation[34]. Considering the time-temperature history of the rheology test, the degree of cure corresponding to the gel time was calculated from the kinetic model for DGEBF/DETDA[35]. An average gel point, across four tests performed at the two different isothermal temperatures, was found to be 0.71, which is in excellent agreement with experimental data found in the literature for the DGEBF/DETDA[34] epoxy system.

Evaluation of post-gelation volumetric shrinkage ($\varepsilon_{sh}$) was performed with the same rheometer setup as the gel point measurement. To measure post-gelation chemical shrinkage (as determined above), the instrument was operated in force-control mode during which a normal force of 0.1 N was applied to maintain contact between the shrinking specimen and the plates. The linear variation ($h - h_0$) in the gap $h$ resulting from the chemical shrinkage in the specimen was continuously monitored by the instrument. The $\varepsilon_{sh}$ resulting from the cure was computed using,

$$\varepsilon_{sh} = \left[1 + \frac{1}{3}\left(\frac{h - h_0}{h_0}\right)\right]^3 - 1 \qquad (1)$$

This relationship assumes that there are no in-plane strains in the specimen and the material is incompressible ($\nu = 0.5$).

The measured $\varepsilon_{sh}$ is shown in Figure 9 for the two different isothermal test temperatures. The average chemical shrinkage in the fully-cured specimens was 2.36 ± 0.08%. With the knowledge of the time-temperature history of the test, the degree of cure was computed using the kinetic model described elsewhere[35]. Figure 10 shows the volumetric shrinkage as a function of degree of cure. It is evident from this figure that the epoxy cures faster at higher temperature, but at the end of the cure-cycle the epoxy ends up fully cured regardless of the processing temperature.

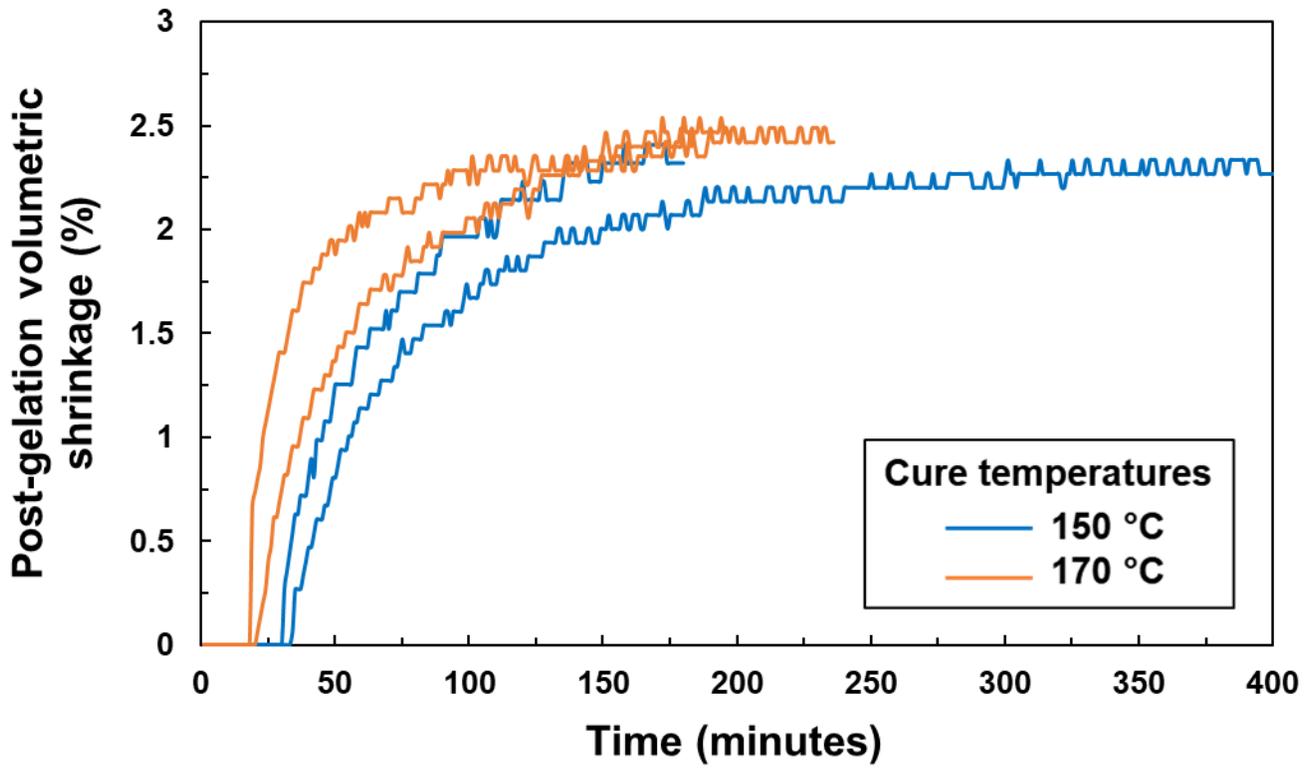

**Figure 9.** Post-gelation volumetric shrinkage as a function of cure time.

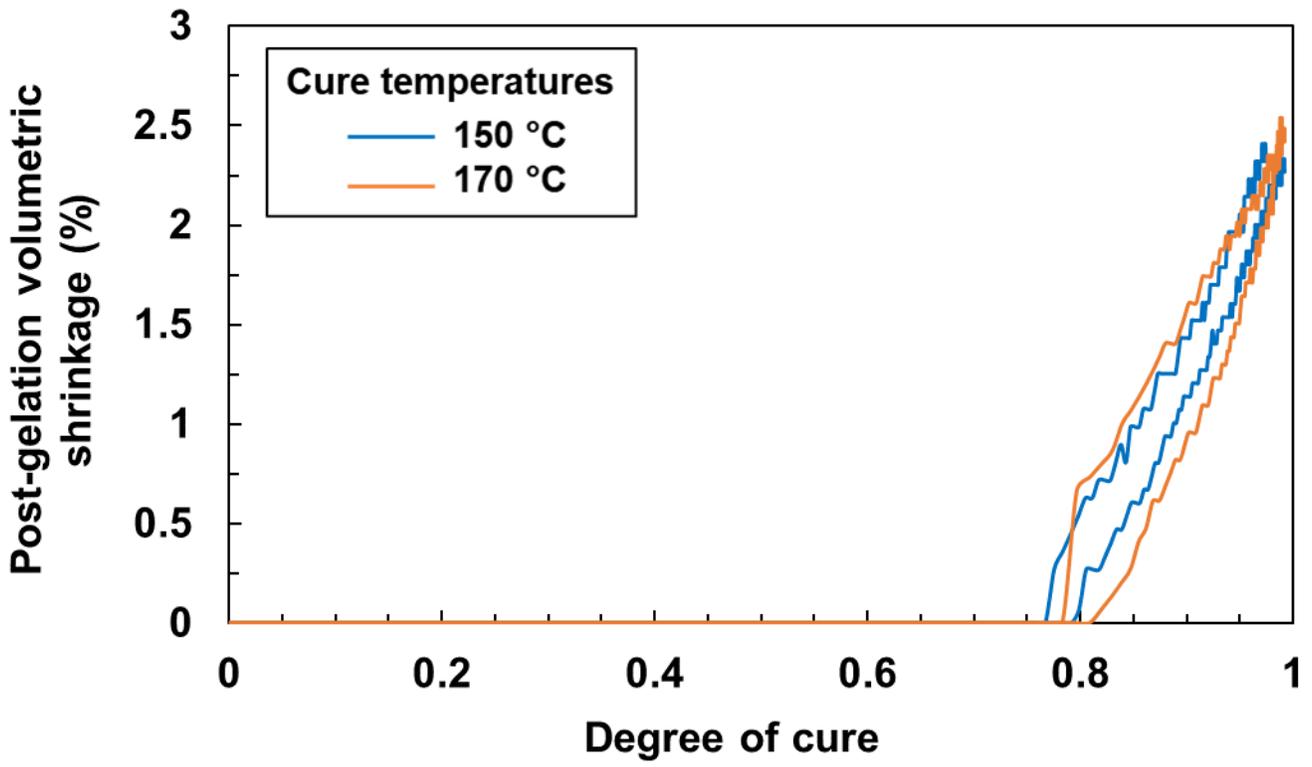

**Figure 10.** Post-gelation volumetric shrinkage as a function of degree of cure.

## 3.3 Mechanical testing

The tensile test specimens were tested according to ASTM D638 guidelines to determine the tensile stiffness, Poisson's ratio, and 0.2% offset yield strength. Each specimen was subjected to a uniaxial tensile load using an MTS Model 43.104 electromechanical test machine in displacement control mode with a crosshead displacement rate of 2 mm/min. During specimen deformation, photographs were captured at even intervals throughout the test up to the point of fracture. A FUJIFILM X-T3 26.1-megapixel camera was utilized for acquiring the photographs.

The open-sourced DIC platform Ncorr [32, 33] built within Matlab was used for the DIC analyses after tensile testing was completed. In performing DIC analyses, deformations of the patterned specimens under loading were measured relative to a reference (undeformed) image through image recognition algorithms[32]. As a result, axial and transverse full-field strains within a user-specified region of interest (ROI) were calculated for each photograph taken during a given test. In this characterization procedure, the full-field strains were averaged for each photograph and then correlated to the corresponding time, load, and displacement readings provided by the MTS apparatus. The ROI was selected to be within the gage area as specified elsewhere[36]. Post-processed data was in the form of stress-strain curves; one for each test specimen.

Characterization of modulus, Poisson's ratio, and offset yield strength was performed using only data from tests in which the specimen fractured within the gage area as designated by ASTM D638. Young's modulus was calculated by performing a fit of the initial linear region of the stress-strain curve up to 1% strain. Poisson's ratio was determined by taking a linear fit of the negative of the strains in the transverse direction divided by the strains the axial direction. The yield strength was measured as the stress corresponding to the strain 0.2% offset from a deviation of the proportional limit of the stress-strain curve. The strain rate was estimated by calculating the average of the slope of adjacent strain-time datapoints.

## 4. Results

This section describes the results of the MD simulations. For each of the plots provided in this section, the black dots represent MD predictions, and the error bars represent the standard error associated with the predictions of the replicate models. The MD data was fitted with appropriate curve fits shown by black dotted trendlines using the OriginPro[37] software. The MD predicted properties for the fully crosslinked model (0.95) are compared with experimentally measured values from the current work and from the literature.

## 4.1 Mass density

Figure 11 shows the mass density as a function of crosslinking density at room temperature. The predicted mass density at 300 K for the fully crosslinked system ($\phi = 0.95$) is found to be $1.207 \pm 0.003$ gm/cm$^3$. This value agrees well with the experimentally-measured value of 1.193 gm/cm$^3$ for fully cured DGEBF/DETDA epoxy in this work. The MD prediction at $\phi = 0.95$ is also compared with experimental value of 1.2 gm/cm$^3$ from the literature[38, 39]. The mass density gradually increases with increasing crosslinking density. This can be attributed to increased network connectivity due to the formation of covalent bonds between the monomers, which reduces the distance between monomers and thus increases the mass density. The mass density data was fitted with a second-order polynomial with $R^2 = 0.998$ as defined by,

$$\rho\left(\frac{gm}{cm^3}\right) = -0.065 \times \phi^2 + 0.142 \times \phi + 1.128 \tag{2}$$

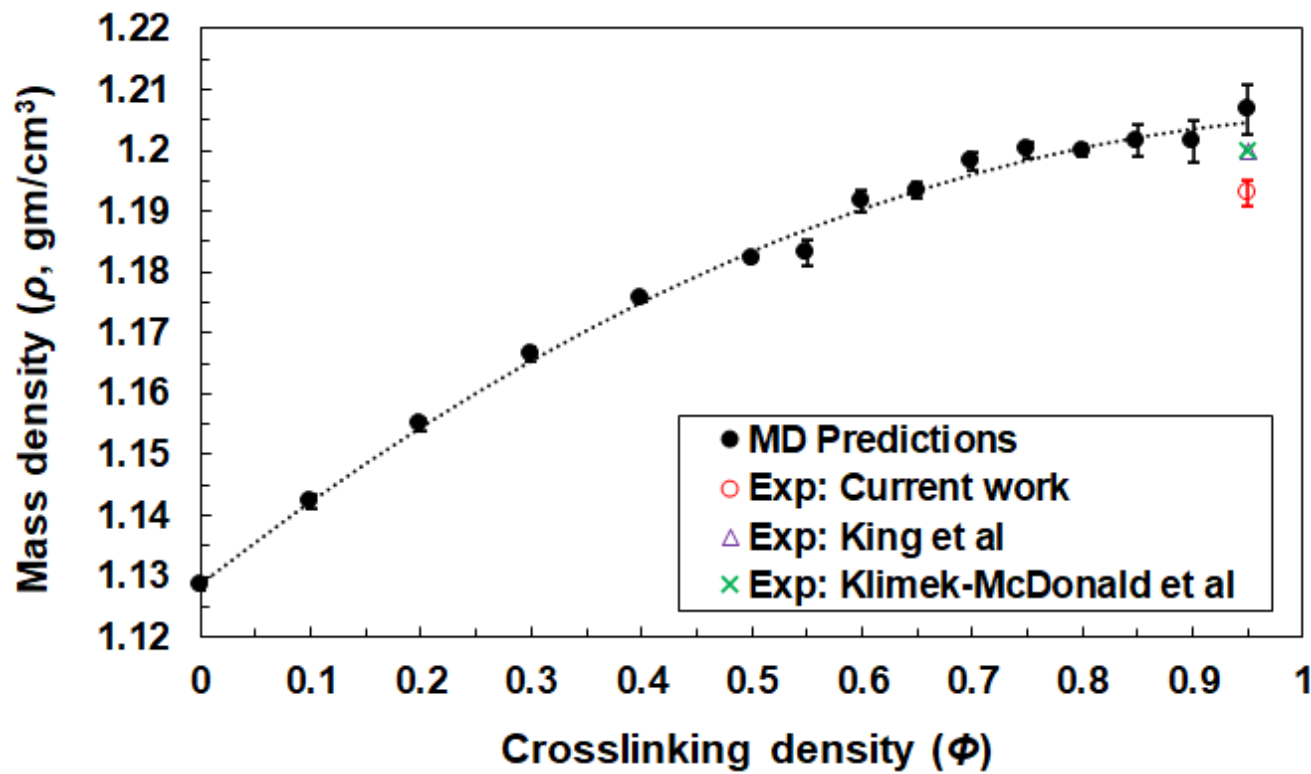

**Figure 11.** Predicted mass density ($\rho$) as a function of crosslinking density at room temperature.

## 4.2   *Volumetric shrinkage*

Figure 12 shows the total volumetric shrinkage as a function of crosslinking density at room temperature. The predicted shrinkage for the fully crosslinked system ($\phi = 0.95$) is 6.496 ±0.184 %. This agrees well with the shrinkage observed during experimental curing of epoxy resins[3]. The volumetric shrinkage gradually increases with increases in crosslinking density because of the formation of covalent bonds. The MD data was fitted using a second-order polynomial with $R^2 = 0.994$ showing a non-linear dependence between volumetric shrinkage and crosslinking density as defined by,

$$Volume\ shrinkage\ (\%) = -5.806 \times \phi^2 + 12.146 \times \phi + 0.044 \tag{3}$$

A similar non-linear trend was observed for this epoxy system using the Dreiding force field[13].

The inset in Figure 12 shows the post-gelation volumetric shrinkage, which is calculated as the change in volume of the crosslinked model with respect to volume of model at $\phi = 0.6$ at the onset of gelation. The MD predicted post-gelation volumetric shrinkage at $\phi = 0.95$ is $1.543 \pm 0.417$ % at room temperature. This value agrees well with the experimentally measured value of $2.36 \pm 0.08$ % in this work. The MD data was fitted using a linear curve fit with $R^2 = 0.912$ showing a linear dependence between post-gelation volumetric shrinkage and crosslinking density as defined by,

$$Post-gelation\ volumetric\ shrinkage\ (\%) = 3.545 \times \phi - 2.003 \qquad (4)$$

This is in general agreement with the experimentally-observed linear dependence between post-gelation shrinkage and degree of cure as shown in Figure 10.

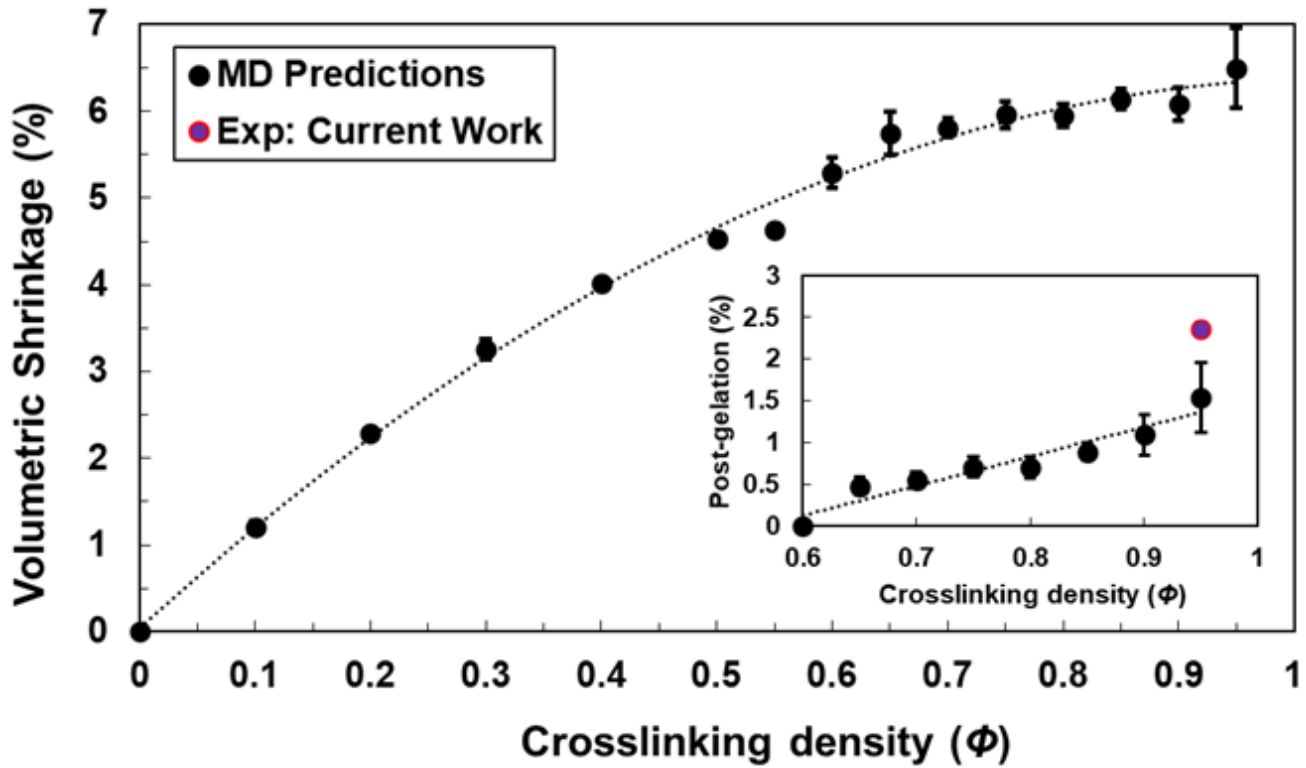

**Figure 12.** Predicted volumetric shrinkage as a function of crosslinking density at room temperature. Post-gelation volumetric shrinkage as a function of crosslinking density at room temperature (inset).

## *4.3  Bulk Modulus*

Figure 13 shows the predicted bulk modulus as a function of crosslinking density. A bulk modulus of $5.546 \pm 0.107$ GPa was predicted for the fully crosslinked system ($\phi = 0.95$). The Bulk modulus gradually increases with crosslinking density, and as it reaches gelation at $\phi = 0.6$, the value approaches a constant value as the material transforms to a solid phase after gelation. Because it is difficult to measure $K$ experimentally, MD predictions are an efficient means of determining this at varying crosslinking densities. The MD data in Figure 13 was fitted with a quadratic curve fit with $R^2 = 0.986$ showing a non-linear dependence between $K$ and crosslinking density as defined by,

$$K\ (GPa) = -1.157 \times \phi^2 + 2.743 \times \phi + 4.283 \tag{5}$$

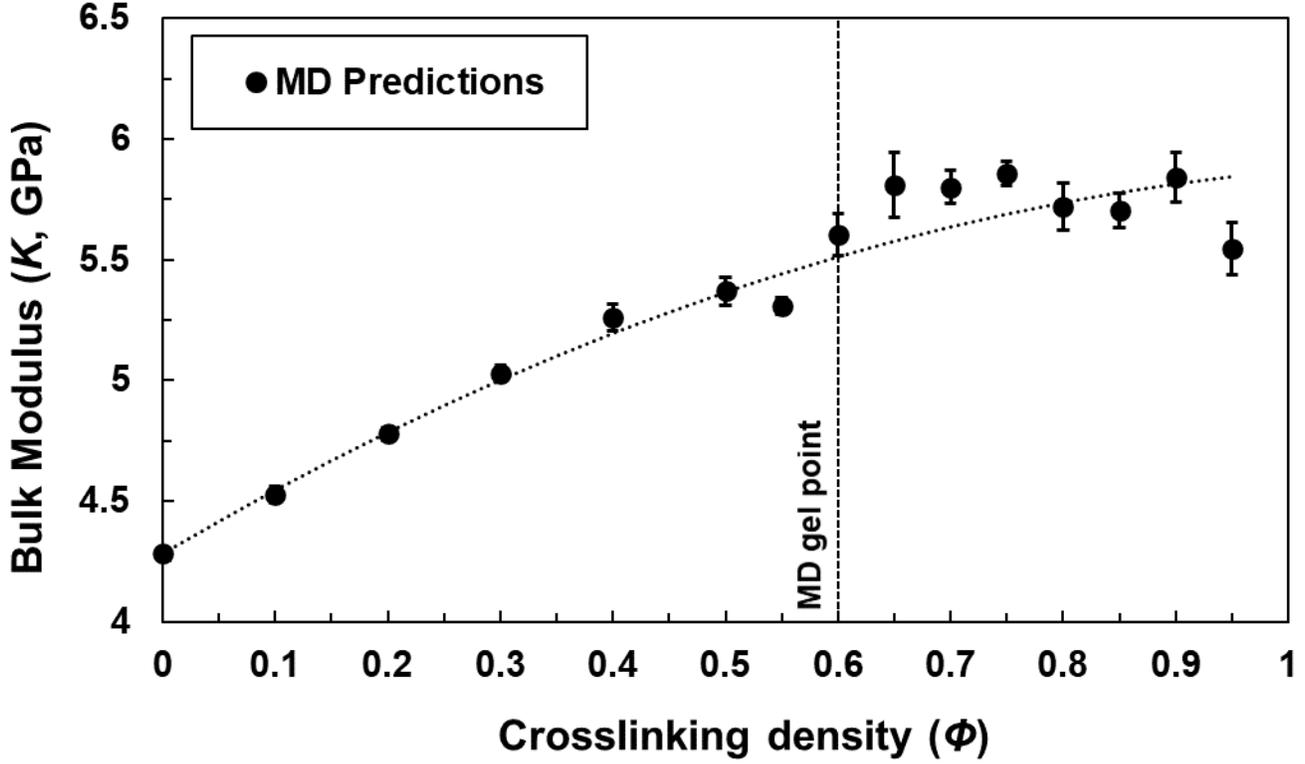

**Figure 13.** Predicted bulk modulus as a function of crosslinking density at room temperature.

## *4.4  Shear Modulus*

Figure 14 shows the predicted shear modulus as a function of crosslinking density. A shear modulus of $1.222 \pm 0.267$ GPa was predicted for $\phi = 0.95$. This value is compared with the experimentally-measured values at three different strain rates from Littell et al[40]. The variation between MD predictions and experimental measurements is due to the orders-of-magnitude difference in strain rates[17]. The shear modulus for the uncrosslinked epoxy is a finite value also because of the strain-rate effect[17]. The MD data in Figure 14 was fitted with a linear curve fit with $R^2 = 0.984$ showing a linear dependence between $G$ and crosslinking density as defined by,

$$G\ (GPa) = 0.969 \times \phi + 0.154 \tag{6}$$

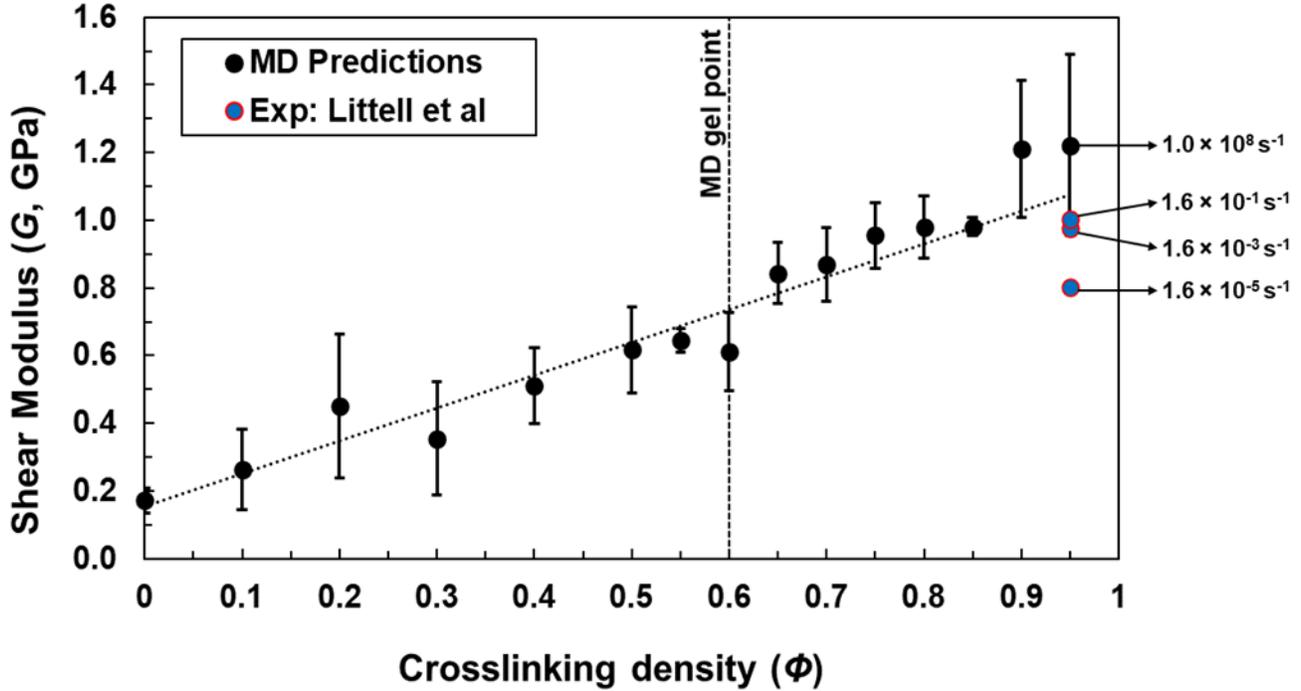

**Figure 14.** Predicted shear modulus as a function of crosslinking density at room temperature. Callouts refer to the corresponding strain rate.

## 4.5 Young's Modulus

Figure 15 shows the predicted Young's modulus as a function of crosslinking density. $E = 3.407 \pm 0.701$ GPa was predicted for $\phi = 0.95$. Also shown in the figure are experimentally-measured values at different strain rates[40, 41]. The experimentally-measured value of 2.392 GPa at $1.95 \times 10^{-4}$ s$^{-1}$ strain rate in this work agrees well with the other experimental data from literature. Despite the strain rate difference between MD simulations and experiments, the predictions at $\phi = 0.95$ match well with higher strain rate ($7 \times 10^2$ s$^{-1}$) experimental data from the Split-Hopkinson Bar test of Gilat et al[41] (results analyzed by Odegard et al[19]). Overall, the Young's modulus gradually increases with increasing crosslinking density. This gradual increase is due to the increased network connectivity, which makes the material stiffer and able to sustain load. No significant change in the magnitude can be seen in the predictions above $\phi = 0.7$, as the material attains gelation. The MD data in Figure 15 was fitted with a linear curve fit with $R^2 = 0.985$ showing a linear dependence between $E$ and crosslinking density as defined by,

$$E \ (GPa) \ = 2.732 \times \phi + 0.461 \qquad (6)$$

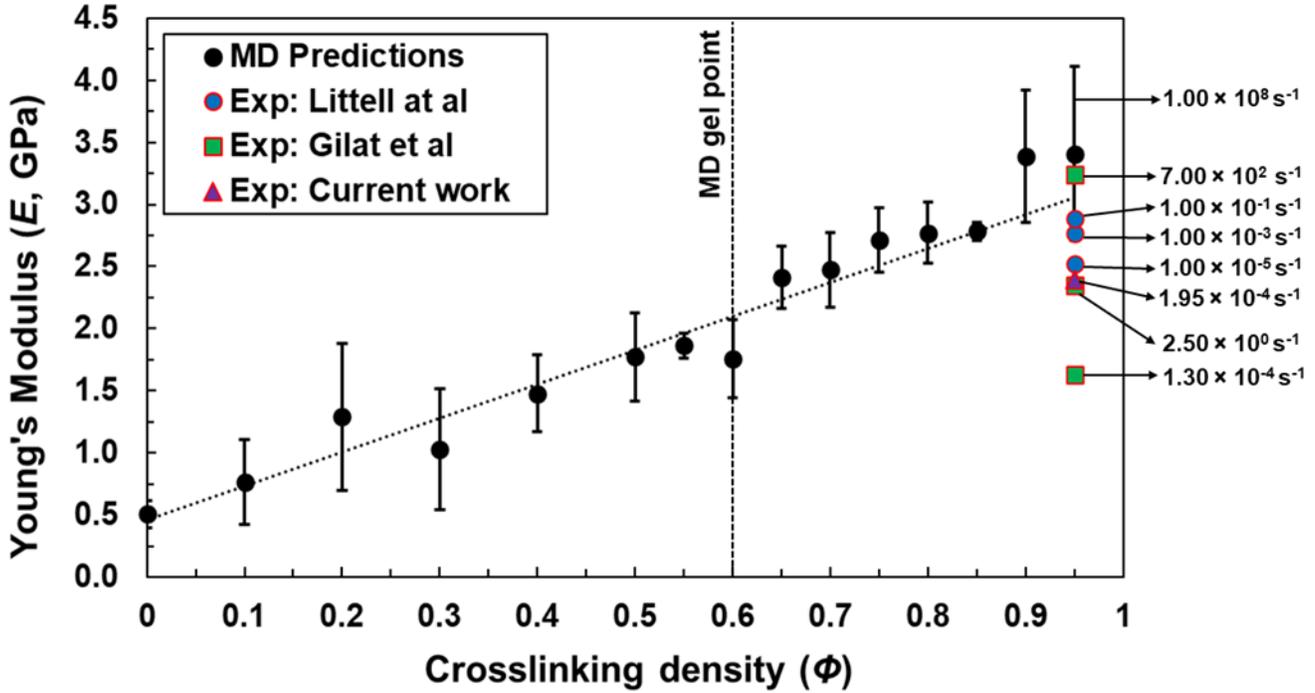

**Figure 15.** Predicted Young's modulus as a function of crosslinking density at room temperature. Callouts refer to the corresponding strain rate.

### 4.6 Poisson's ratio

Figure 16 shows the predicted Poisson's ratio as a function of crosslinking density. A Poisson's ratio of $0.398 \pm 0.019$ was predicted for $\phi = 0.95$ and $0.480 \pm 0.004$ was predicted for $\phi = 0$. Also shown in the figure are the experimentally-measured values at different strain rates from Littell et al[40]. The experimentally-measured value of 0.366 at $1.95 \times 10^{-4}\,\text{s}^{-1}$ strain-rate in this work agrees well with the data from the literature. The Poisson's ratio gradually decreases with increasing crosslinking density. The MD predictions at $\phi = 0.95$ agree well with the experimental values despite the difference in strain rates, which demonstrates a smaller influence of the strain rate on Poisson's ratio relative to Young's modulus and shear modulus. The MD data in Figure 16 was fitted with a quadratic curve fit with $R^2 = 0.985$ showing a non-linear dependence between $v$ and crosslinking density as defined by,

$$v = -0.012 \times \phi^2 - 0.063 \times \phi + 0.480 \tag{7}$$

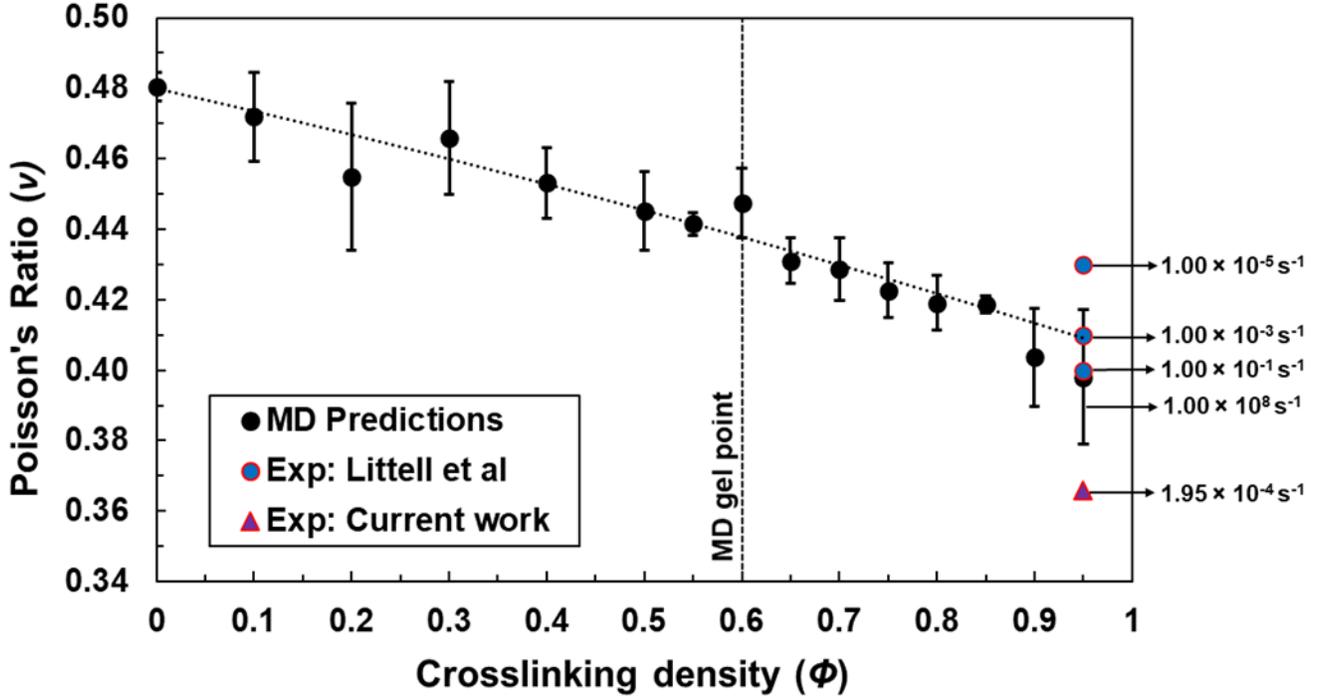

**Figure 16.** Predicted Poisson's ratio as a function of crosslinking density at room temperature. Callouts refer to the corresponding strain rate.

## *4.7 Yield strength*

Figure 17 shows the predicted yield strength as a function of crosslinking density. A yield strength of 107.67 ± 16.67 MPa was predicted for $\phi = 0.95$. In the figure, this value is compared with experimentally-measured values at different strain rates[40, 41]. The experimentally-measured value of 36.96 MPa at a $1.95 \times 10^{-4}$ s$^{-1}$ strain rate in this work agrees well with the experimental data from the literature. It can also be seen in Figure 17 that at $\phi = 0.95$ the yield strength increases with increasing strain rate and the experimental value at the highest strain rate from Gilat et al[41] (as analyzed by Odegard et al[19]) matches well with the MD prediction. The yield strength increases with increasing crosslinking density due to increased network connectivity. In the pre-gelation regime ($\phi = 0$ to 0.6) the yield strength is low, as the material is a viscous liquid and cannot sustain large mechanical loads. The yield strength value increases quickly as the material reaches gelation ($\phi = 0.6$) and continues to increase in the post-gelation regime ($\phi = 0.6$ to 0.95) as the material attains a tighter network and can sustain significant loads. The MD data is fitted with a sigmoidal (Boltzmann equation) curve fit with $R^2 = 0.885$ which correctly represents the evolution of the yield strength of the epoxy from the un-crosslinked liquid phase to the fully crosslinked solid phase as defined by,

$$\sigma\ (MPa) = 123.839 + \frac{(0.983 - 123.839)}{(1 + e^{\left(\frac{\phi - 0.76}{0.1}\right)})} \tag{8}$$

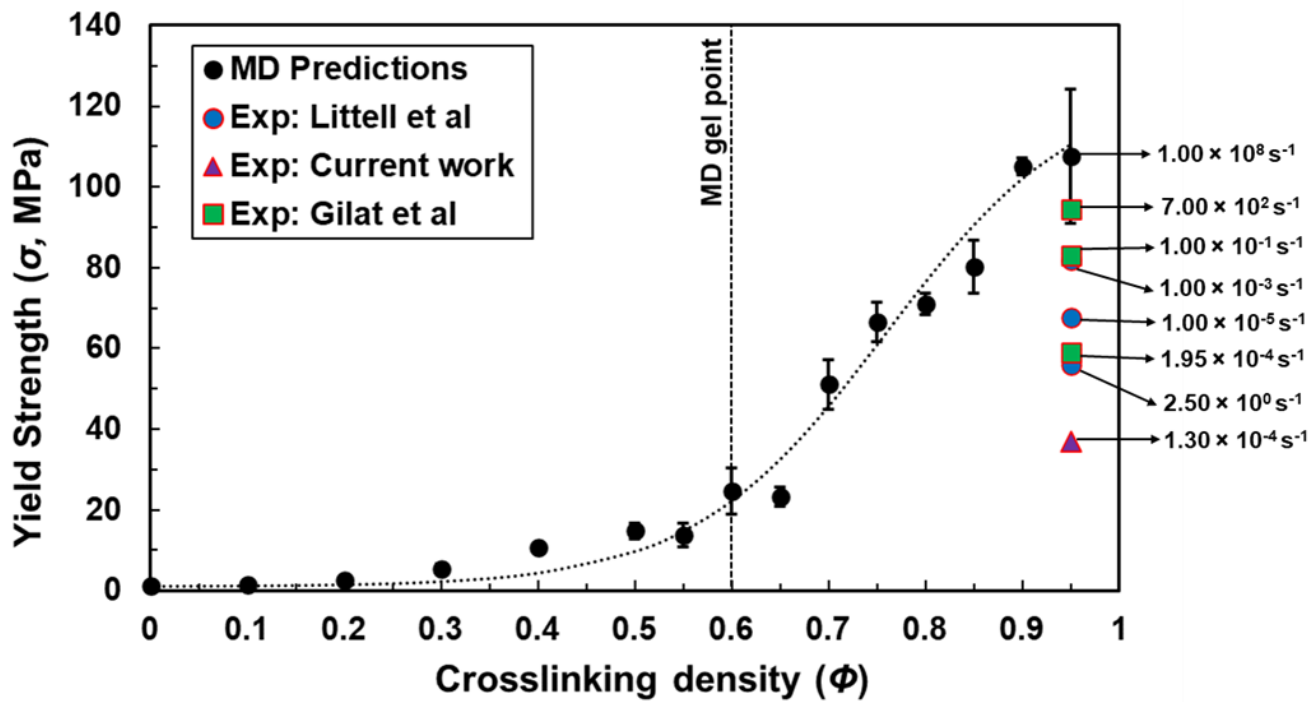

**Figure 17.** Predicted yield strength as a function of crosslinking density at room temperature. Callouts refer to the corresponding strain rate.

## 5. Conclusions

The computational predictions from this study show that the epoxy mass density, volumetric shrinkage, stiffness, strength, and Poisson's ratio are strongly affected by the crosslink density, especially below the gel point. Therefore, each of these properties is highly dependent on the extent of the polymer network up to the point at which the network extends indefinitely in the polymer and can thus bear a mechanical load and resist deformation. The predictions generally agree with the experimental measurements performed herein as well as those from the literature. Therefore, the computational simulation protocols are effective, and IFF-R is reliably predicting accurate physical and mechanical properties. The predicted properties at varying crosslinking densities provide insight into the evolution of properties of the epoxy system during the processing of composite materials. This data can be used as input into higher length-scale computational tools to predict the residual stresses that are developed during the processing of these complex networked epoxies in the presence of various composite reinforcements. Such modeling can be used to optimize processing parameters and improve composite laminate strength and reduce post-manufacturing residual deformations.

## Supporting Information

## Acknowledgement

This research was partially supported by the NASA Space Technology Research Institute (STRI) for Ultra-Strong Composites by Computational Design (US-COMP), grant NNX17AJ32G; and NASA grant 80NSSC19K1246. SUPERIOR, a high-performance computing cluster at Michigan Technological University, was used in obtaining the MD simulation results presented in this publication.

## Data Availability

The raw/processed data required to reproduce these findings cannot be shared at this time as the data also forms part of an ongoing study.